\documentclass[fleqn,usenatbib]{mnras}

\usepackage{newtxtext,newtxmath}
\usepackage[normalem]{ulem}

\usepackage[T1]{fontenc}

\DeclareRobustCommand{\VAN}[3]{#2}
\let\VANthebibliography\thebibliography
\def\thebibliography{\DeclareRobustCommand{\VAN}[3]{##3}\VANthebibliography}

\usepackage{graphicx}	
\usepackage{amsmath}	

\title[Breaking the Blend in NGC~5728]{Breaking the Blend: A multi-tracer kinematic decomposition method for IFS data applied to disentangling the AGN outflow and circumnuclear ring in NGC~5728 with JWST}

\author[O. Veenema et al.]{
Oscar Veenema,$^{1}$\thanks{E-mail: oscar.veenema@physics.ox.ac.uk}
Niranjan Thatte,$^{1}$
Dimitra Rigopoulou,$^{1,2}$
Richard I. Davies,$^{3}$
Miguel Pereira-Santaella,$^{4}$
\newauthor
Ismael Garc\'{i}a-Bernete,$^{5}$
Almudena Alonso-Herrero,$^{5}$
Anelise Audibert,$^{6, 7}$
Enrica Bellocchi,$^{8, 9}$
\newauthor
Andrew J. Bunker,$^{1}$
Fran\c{c}oise Combes,$^{10}$
Tanio D\'{\i}az Santos,$^{11, 2}$
Fergus R. Donnan,$^{12}$
\newauthor
Donaji Esparza-Arredondo,$^{13}$
Federico Esposito,$^{14}$
Santiago Garc\'{i}a-Burillo,$^{14}$
Bego{\~n}a Garc\'{i}a-Lorenzo,$^{6,7}$
\newauthor
Omaira Gonzalez Martin,$^{15}$
Laura Hermosa Mu{\~n}oz,$^{16}$
Erin K. S. Hicks,$^{17}$
Sebastian F. H\"onig,$^{18}$
\newauthor
Masatoshi Imanishi,$^{19, 20}$
Alvaro Labiano,$^{21}$
Nancy A. Levenson,$^{22}$
Enrique Lopez-Rodriguez,$^{23}$
\newauthor
Cristina Ramos Almeida,$^{6, 7}$
Claudio Ricci,$^{24, 25}$
Rogemar A. Riffel,$^{26}$
Marko Stalevski$^{27, 28}$
\\
\\
\emph{\normalsize Affiliations are listed at the end of the paper}
}

\date{Accepted XXX. Received YYY; in original form ZZZ}

\pubyear{\the\year{}}

\begin{document}
\label{firstpage}
\pagerange{\pageref{firstpage}--\pageref{lastpage}}
\maketitle

\begin{abstract}

Integral field spectroscopy (IFS) of the central kiloparsecs of active galactic nuclei (AGN) reveals a mixture of spatially coincident emission from star-formation and AGN feedback exciting the interstellar medium. Disentangling these components remains a challenge, as most spectral tracers are affected by both processes, limiting robust interpretation of kinematics and energetics. We present a new framework for decomposing IFS data into distinct components on a spaxel-by-spaxel basis using a multi-tracer, stacked kinematics approach. This method combines kinematic modelling with imposed flux decomposition per spaxel, quantifying the contribution of each component across the field of view. We apply this method to JWST IFS observations of the Seyfert galaxy NGC~5728 from the Galaxy Activity, Torus, and Outflow Survey (GATOS), analysing twelve mid-infrared fine-structure lines ($4.49 \leq \lambda \leq 25.89 \mu$m, $7.9 \leq$ IP $\leq 126.2$ eV). We find that the circumnuclear emission can be decomposed into two dominant components: a star-forming ring and an AGN-driven biconical outflow. Our method separates these structures and recovers their detailed spatial morphology. This framework provides a general and scalable method for physically motivated component separation in IFS data, applicable across many wavelength ranges and targets, enabling reliable interpretation of complex emission line structures (or morphologies) in active galaxies and beyond.

\end{abstract}




\section{Introduction}
\label{sec:Introduction}

Integral field spectroscopy (IFS) has revolutionised the study of galaxies by providing spatially resolved spectroscopy of extended sources. In active galactic nuclei (AGN), however, interpreting these observations remains challenging because emission from AGN feedback and star-formation is often superposed within individual spaxels. Consequently, the observed spectrum is a composite of emission from multiple physical mechanisms (components), making it difficult to disentangle and recover the true morphology and energetics of AGN activity and star-formation (e.g., \citealt{DaviesRL2016, Lai2022, HermosaMunoz2026}). Existing approaches typically rely on individual emission line measurements or diagnostics to identify these different components (e.g., \citealt{davies2014starburst, Feltre2016, Kirkpatrick2017, kewley2019understanding, DAgostino2019, law2021sdss, zhang2025theoretical}), including through modelling kinematics (e.g., \citealt{marconcini2025fast, ceci2026miracle, veenema2026kinematics, marconcini2026miracle, Donnan2026}). However, because emission from individual optically thin lines generally arises from multiple overlapping structures along the line of sight, interpreting these diagnostics can be challenging.

The James Webb Space Telescope (JWST) integral field spectrographs (IFSs), namely the Near-Infrared Spectrograph (NIRSpec) \citep{Jakobsen_2022, boker2022near} and Mid-Infrared Instrument/Medium Resolution Spectrometer (MIRI/MRS) \citep{wells2015mid, argyriou2023jwst}, have transformed studies of nuclear and circumnuclear regions of active galaxies through combination of sub-arcsecond resolution and exceptional sensitivity. Such observations have revealed a rich interplay between circumnuclear star-forming rings and discs (e.g., \citealt{bianchin2024goals}), AGN-driven outflows (e.g., \citealt{zhang2024galaxy, munoz2024biconical, ulivi2025jwst}), shocked gas, and other structures that influence the physical conditions of the nuclear interstellar medium (ISM) (e.g., \citealt{davies2024gatos, veenema2025shock, almeida2025jwst, Riffel2026, Donnan2026}). However, this increased complexity makes physically motivated component separation a central challenge for interpreting modern IFS observations.

In this Letter, we present a novel decomposition framework that disentangles, quantifies, and spatially maps physically distinct emission components in IFS data on a spaxel-by-spaxel basis. Because the method is independent of specific emission line tracers and observing facilities, it can be applied consistently across a wide range of wavelength regimes and integral field spectrographs. We demonstrate the method using JWST IFS observations of NGC~5728, a nearby Seyfert~2 galaxy hosting a prominent circumnuclear star-forming ring and an AGN-driven biconical outflow. Our framework separates these components, enabling their emission to be spatially mapped and analysed independently. We include additional figures and information in the supplementary material.

\section{Target, observations and data reduction}
\label{sec:Data collection}

NGC~5728 is a nearby Seyfert~2 galaxy ($z = 0.00932$, $D \sim 39$ Mpc; \citealt{shimizu2019multiphase}, thus 1 arcsecond $\sim 190$ pc), hosting a Compton-thick AGN with a black hole mass of $3.4 \times 10^{7}~M_\odot$ \citep{durre2019agn}. It exhibits a $\sim 1.5$ kpc-scale biconical ionised outflow \citep{wilson1993, Durre2018, davies2024gatos} that intersects a prominent circumnuclear star-forming ring (with a radius of $\sim 800$ pc; \citealt{shimizu2019multiphase, Shin2019, Falcao2024}), making it an ideal target for applying such a decomposition method, with strong emission from both of these components.

We analyse combined JWST/NIRSpec and MIRI/MRS observations of NGC~5728 spanning $\sim2.7$-$27.9 \mu$m and probing the inner $\sim1.5$ kpc of the galaxy. These data (NIRSpec IFU: PI: Garc\'{i}a-Bernete; programme ID: 5017, MIRI/MRS: PI: T. Shimizu and R. Davies; programme ID: 1670) are drawn from the Galaxy Activity, Torus, and Outflow Survey (GATOS) \citep{garcia2021galaxy, alonso2021galaxy, garcia2024galaxy} and were originally presented by \citet{Donnan2026} for NIRSpec/IFU and \citet{davies2024gatos} for MIRI/MRS, to which we refer the reader for further details of the observations and data reduction.

\section{Method and Results}
\label{sec:method_and_results}

Our method requires IFS data containing spectral features that trace the physical components of interest. Emission lines are well suited. Distinct physical components can produce the same lines, but with different kinematic signatures, which we exploit to disentangle them.

\subsection{Fitting emission line kinematics}

\begin{figure}
   \centering
   \includegraphics[width=\columnwidth]{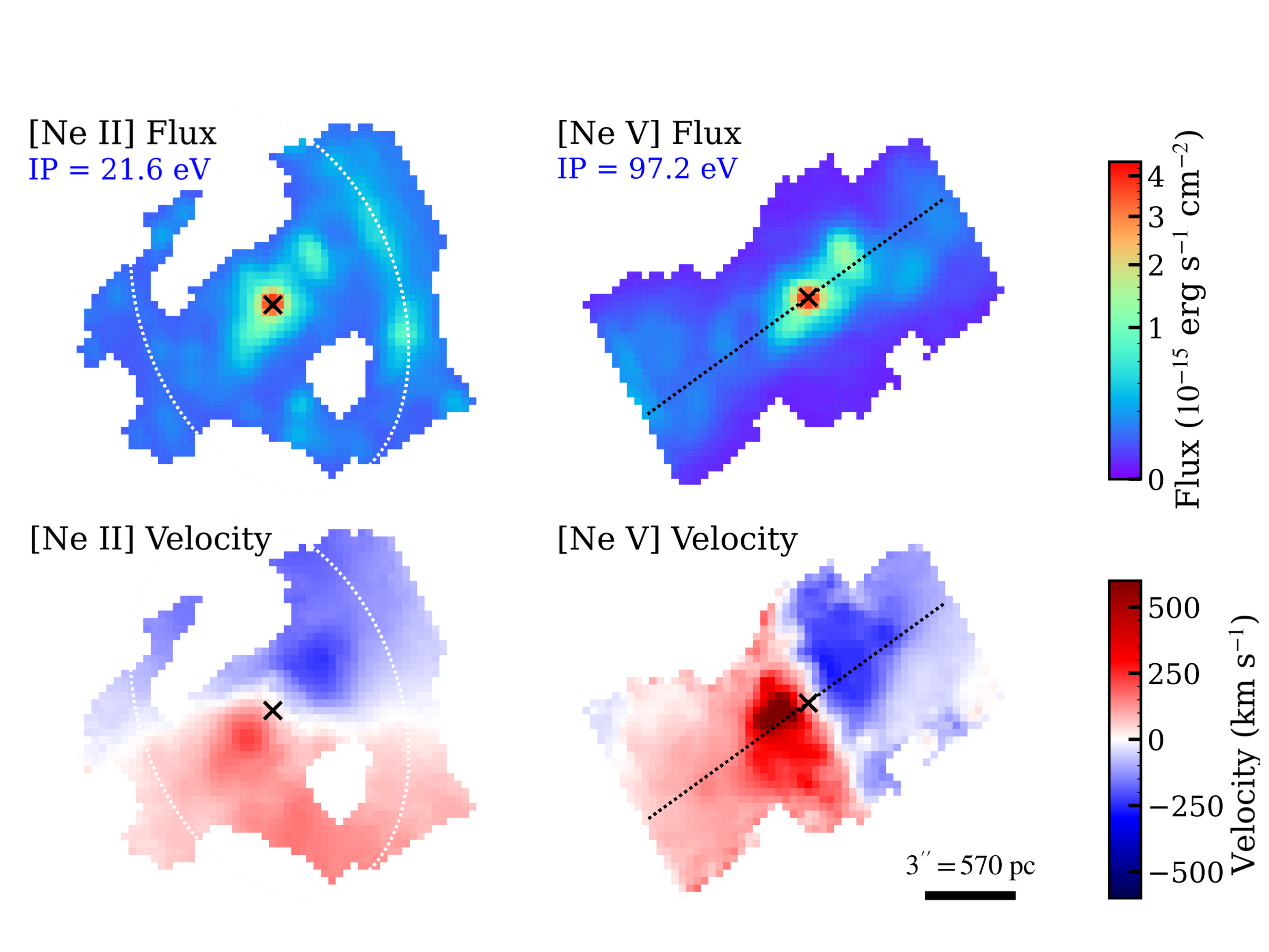}
   \caption{Integrated flux maps (top) and single-Gaussian fit velocity maps (bottom) of the low-IP [Ne II] 12.81 $\mu$m line (left) and high-IP [Ne V] 14.32 $\mu$m line (right), illustrating the circumnuclear morphology and kinematics of NGC~5728. The star-forming ring and the major axis of the biconical outflow are indicated by the white dotted ellipse and black dotted line, respectively. The [Ne II] flux and velocity maps trace both the star-forming ring and the AGN-driven outflow, while [Ne V] is confined to the AGN and its outflow. North is up and east is to the left. The cross denotes the AGN position.}
    \label{fig:NeII_NeV_comparison}%
\end{figure}

For each emission line and spaxel, we masked the line profile and fitted a third-order polynomial to the continuum over a 0.02~$\mu$m interval on each side, which was then subtracted to isolate the line emission. We excluded spaxels with line fluxes below the 60th percentile for that transition, retaining the brightest 40\% of spaxels. This preferentially selects regions with sufficient S/N for reliable Gaussian kinematic fitting, as discussed by \citet{veenema2026kinematics}. The resulting minimum S/N exceeds 6 for all lines, providing a balance between spatial coverage and the statistical robustness necessary for clearly identifying multiple Gaussian components within the remaining spaxels \citep{garcia2013searching}. For each selected spaxel, we fitted both single- and double-Gaussian profiles to every emission line\footnote{We do not correct the measured velocity dispersions for the wavelength-dependent instrumental line-spread function, as our goal is not to recover their intrinsic values, but rather to constrain the broadening of each component within each IFS channel and thereby separate their contributions to the observed line profile. Importantly, although the line spread function varies between channels, this does not bias the subsequent component decomposition, as our stacking method uses the integrated line flux, for which the increase in velocity dispersion is compensated by a corresponding decrease in amplitude.}. All Gaussian fits were performed using Levenberg-Marquardt optimisation implemented in the CapFit Python library \citep{cappellari2023full}. Fig.~(\ref{fig:NeII_NeV_comparison}) shows the integrated flux and velocity maps of the [Ne II] 12.81 $\mu$m and [Ne V] 14.32 $\mu$m emission lines, revealing the distinct signatures of the disc and outflow.

\subsection{Kinematic component modelling}

Our initial aim is to quantify the key kinematic parameters of the dominant components exciting emission lines in the IFS data. In particular, we require reliable estimates of the line of sight velocity and velocity dispersion of each component in each spaxel. These can then be used in subsequent fits, allowing us to decompose the emission flux from each spaxel into multiple components.

Ionic lines from heavy elements trace gas across a range of ionisation states of the ISM and are sensitive to photoionisation from stars and AGN, as well as their associated feedback processes. In this work we focus on such ionic lines in NGC~5728 in the mid-IR, enabling us to probe gas over a wide range of ionisation potentials (IPs) and thereby trace both stellar and AGN activity. The ionic lines analysed in our JWST data, together with their wavelengths and IPs, are listed in Table~\ref{tab:ion_lines}. These were selected for their strong spaxel-by-spaxel detections, relatively clean spectral regions, and broad IP coverage. While our method benefits from multiple transitions, it remains applicable with fewer or alternative lines.

\begin{table*}
    \centering
    \caption[]{Mid-IR emission lines analysed in the JWST IFS data of NGC~5728, listing ionisation potential (IP; the energy required to reach the ionisation stage producing each transition; \citealt{kramidateam}), rest-frame wavelength, and median velocity dispersions for the disc and outflow components ($\sigma_{v-\rm disc}$ and $\sigma_{v-\rm outflow}$) derived from fully unconstrained double-Gaussian fits across all spaxels, with uncertainties given by the 16th-84th percentile range. Also listed is the mean disc flux fraction, $f_{\rm disc}$ (with $f_{\rm outflow} = 1 - f_{\rm disc}$), together with the statistical weight assigned to each line in the stacked $f_{\rm disc}$ calculation in Section~\ref{sec:stacked_approach}, based on its average $f_{\rm disc}$. The [Ne V] and [Ne VI] lines show no detectable disc component in these data of NGC~5728}, thus we can use them to constrain the outflow. Conversely, no ionic lines trace purely the disc.
    \label{tab:ion_lines}
    \begin{tabular}{lcccccc}
        \hline
        Emission Line & IP (eV) & Wavelength ($\mu$m) & Median $\sigma_{v-\rm disc}$ (km s$^{-1}$) & Median $\sigma_{v-\rm outflow}$ (km s$^{-1}$) & Average $f_{\rm disc}$ & Statistical weight \\
        \hline
        {[Fe II]}  & 7.9  & 5.34  & $78^{+49}_{-36}$  & $203^{+129}_{-77}$ & 0.45 & 0.90 \\
        {[Ar II]}  & 15.8 & 6.99  & $59^{+31}_{-14}$  & $159^{+46}_{-58}$  & 0.57 & 0.86 \\
        {[Ne II]}  & 21.6 & 12.81 & $55^{+45}_{-10}$  & $134^{+19}_{-28}$  & 0.60 & 0.81 \\
        {[S III]}  & 23.3 & 18.71 & $72^{+27}_{-8}$   & $175^{+75}_{-66}$  & 0.55 & 0.90 \\
        {[Ar III]} & 27.6 & 8.99  & $67^{+33}_{-15}$  & $163^{+52}_{-58}$  & 0.57 & 0.87 \\
        {[S IV]}   & 34.9 & 10.51 & $80^{+34}_{-23}$  & $193^{+44}_{-52}$  & 0.47 & 0.95 \\
        {[Ne III]} & 41.0 & 15.56 & $72^{+36}_{-13}$  & $194^{+42}_{-47}$  & 0.33 & 0.67 \\
        {[O IV]}   & 54.9 & 25.89 & $81^{+52}_{-17}$  & $176^{+77}_{-56}$  & 0.41 & 0.81 \\
        {[Ar VI]}  & 74.8 & 4.53  & $51^{+36}_{-34}$  & $139^{+58}_{-57}$  & 0.29 & 0.59 \\
        {[Mg IV]}  & 80.1 & 4.49  & $71^{+63}_{-41}$  & $197^{+86}_{-79}$  & 0.25 & 0.49 \\
        {[Ne V]}  & 97.2 & 14.32  & -  & $255^{+149}_{-91}$  & $\sim0$ & 0 \\
        {[Ne VI]}  & 126.2 & 7.65  & -  & $150^{+48}_{-50}$  & $\sim0$ & 0 \\
        \hline
    \end{tabular}
\end{table*}

For ionic species, IP is the primary discriminator of ionisation origin: low-IP lines (IP $\lesssim 20$ eV) predominantly trace star-forming regions, whereas high-IP lines (IP $ > 55$ eV) mainly trace AGN photoionisation and feedback, including AGN driven outflows \citep{munoz2024biconical, veenema2026kinematics}. Additional line-specific effects further enhance their diagnostic power. For example, [Fe II] 5.34 $\mu$m can trace shock-heated gas, as shocks liberate iron from dust grains \citep{kawara1988forbidden, colina2015understanding, vivian2022goals, herrero2025miconic}, and [Mg IV] 4.49 $\mu$m may trace faster shocks via similar processes \citep{pereira2024extended}. Although [S IV] 10.51 $\mu$m has an intermediate-IP, it lies within the 9.8 $\mu$m silicate absorption feature, making it especially sensitive to dust attenuation \citep{zhang2024galaxy}. [O IV] 25.89 $\mu$m, with an IP comparable to He II, can be ionised by both AGN activity and hot O-type stars \citep{thornley2000massive, pereira2010mid, AlonsoHerrero2012}. Collectively, these effects make mid-IR ionic lines powerful multi-phase tracers of the ISM, often extending beyond a simple dependence on IP. Thus, any individual emission line provides a biased representation of the component being studied, motivating the use and combination of multiple emission line tracers.

We find that all emission lines in Table~\ref{tab:ion_lines} trace at least one, and in most cases both, of the circumnuclear star-forming ring/disc and the AGN driven biconical outflow in NGC~5728 (See figures~S1 and S2 in the supplementary material for the integrated line flux and single-Gaussian velocity map of each line). The [Ne V] and [Ne VI] lines are found to almost exclusively trace the AGN-driven outflow in NGC~5728. This is consistent with their very high IPs, but does not represent a global trend for all AGN, as the result depends strongly on the coupling between the disc and outflow and thus the ability of AGN radiation to excite gas in the disc. Disc and outflow components often produce overall emission line profiles that are well described by a double-Gaussian function. The outflow component typically exhibits broader profiles than the star-forming disc, reflecting the higher velocity dispersion of gas in the bicone compared to the more ordered disc rotation \citep{munoz2024biconical, veenema2026kinematics}. By enforcing $\sigma_{\rm outflow} > \sigma_{\rm disc}$ in each spaxel, we derive the distribution of velocity dispersions associated with each component for each emission line across all spaxels in which they are well detected, as shown in figure~S3 of the supplementary material.

Although assuming $\sigma_{\rm outflow} > \sigma_{\rm disc}$ may not hold universally, the resulting dispersion distributions are well constrained across hundreds of spaxels, providing robust estimates for the decomposition. Accurate velocity field modelling is also essential, and our framework can be applied to any component given an appropriate velocity model. For NGC~5728, we adopt a thin inclined rotating disc model for the circumnuclear ring and a hollow bicone model for the ionised AGN outflow. The parameters for each component must either be fitted from the data or adopted from the literature.

We emphasise that our method is not tied to any particular kinematic model; models can be chosen with the complexity required by the system under study. For example, \texttt{3DBarolo} \citep{teodoro20153d} can model discs with bars or warps, while \texttt{MOKA3D} \citep{marconcini2023moka3d} can describe outflows with more complex morphologies. For the JWST data considered here, where the disc is only partially covered by the ionised gas emission, we find that simple thin rotating disc and hollow bicone models provide excellent fits, with residuals typically below the instrumental dispersion of the relevant IFU channel. For completeness, we discuss fits using \texttt{3DBarolo} and \texttt{MOKA3D} in the supplementary material.

\subsubsection{Thin inclined rotating disc}
\label{sec:disc_model}

We firstly consider the circumnuclear ring, for which we find a simple thin inclined rotating disc model adequately reproduces the observed velocity field despite the known nuclear bar \citep{shimizu2019multiphase}. This model (also applied by \citealt{veenema2026kinematics, Donnan2026}), describes the projected two-dimensional velocity field of a thin rotating disc, with sky coordinates $(x, y)$ defined relative to the galaxy centre. We adopt a $\tanh$ rotation curve to parameterise the line-of-sight velocity field, capturing the characteristic central rise and outer flattening typical of disc rotation, and providing an empirical fit to the data. The resulting disc velocity field is given by:

\begin{equation}
v_{\text{disc-model}}(x, y) = v_{\text{sys}} + v_{\text{disc}} \tanh\left(\frac{R_{\text{}}}{R_{\text{disc}}}\right) \sin(i_{\rm disc}) \cos(\theta_{\rm disc}) .
\end{equation}

Here, $v_{\rm disc}$ (maximum disc velocity) and $R_{\rm disc}$ (disc turnover radius) are fitting parameters, and $v_{\text{sys}}$ is the systemic velocity of the galaxy. In this model, $R$ is the deprojected disc-plane radius and $\theta_{\rm disc}$ the corresponding azimuthal angle, defined as:

\begin{align}
R &= \sqrt{x'^2 + \left( \frac{y'}{\cos i_{\rm disc}} \right)^2} \\
\theta_{\rm disc} &= \arctan\left( \frac{y'}{\cos i_{\rm disc} \cdot x'} \right) \\
x' &= x \cos\phi_{\rm disc} + y \sin\phi_{\rm disc} \\
y' &= -x \sin\phi_{\rm disc} + y \cos\phi_{\rm disc} .
\end{align}

where $\phi_{\rm disc}$ is the position angle (PA) of the kinematic major axis, and $(x', y')$ are the rotated $(x, y)$ coordinates aligned with the major axis. The circumnuclear disc has been shown to have an inclination of $i_{\rm disc} \sim 43^\circ$ and a PA of $\phi_{\rm disc} \sim 194^\circ$ east from north \citep{shimizu2019multiphase}. As our JWST observations do not fully cover the circumnuclear ring, we adopt these literature values for our disc model. Nevertheless, fitting the low-IP ionic lines ([Fe II], [Ar II], and [Ne II]) with our thin inclined rotating disc model, and \texttt{3DBarolo} \citep{teodoro20153d}, both yield values consistent with the literature, providing an independent check of their applicability to the ionised gas. We therefore fix $i_{\rm disc}$ and $\phi_{\rm disc}$ and fit only for $v_{\rm disc}$ and $R_{\rm disc}$.

\subsubsection{AGN-driven biconical outflow}
\label{sec:bicone_model}

The second kinematic component is the AGN-driven outflow, which we model as an inclined hollow bicone inspired by the framework of \citet{das2005mapping, das2006kinematics}. However, rather than constructing a full three-dimensional velocity cube, we directly evaluate the projected velocity field in the plane of the sky, retaining the hollow bicone geometry and a similar piecewise linear radial velocity law. In this picture, gas is launched radially within a hollow double cone, with emission confined to an angular range about the outflow axis. This simplified description provides an excellent fit to the data while remaining computationally efficient for Markov Chain Monte Carlo (MCMC) optimisation (see Section.~\ref{sec:MCMC}). The model is parameterised by the outflow inclination ($i_{\rm outflow}$), position angle ($\phi_{\rm outflow}$), opening angles, and a radial velocity law as a function of distance from the AGN.

We again adopt a coordinate system centred on the AGN, with the same projected sky-plane coordinates $(x, y)$ and line-of-sight direction along $z$. The outflow axis is defined by $i_{\rm outflow}$ and $\phi_{\rm outflow}$, giving the unit vector:

\begin{equation}
\hat{\mathbf{n}} =
(\sin i_{\rm outflow}\cos{\rm \phi_{\rm outflow}}, \\
\sin i_{\rm outflow}\sin{\rm \phi_{\rm outflow}}, \\
\cos i_{\rm outflow}).
\end{equation}

For each spaxel, the unit position vector is
\begin{equation}
\hat{\mathbf{r}} = \frac{(x, y, 0)}{R^*}, \quad R^* = \sqrt{x^2 + y^2},
\end{equation}
so that the angle between the projected radial position vector and the outflow axis is $\cos\theta_{\rm outflow} = \hat{\mathbf{r}} \cdot \hat{\mathbf{n}}$. Emission is restricted to a hollow bicone defined by inner and outer opening angles, $\theta_{\rm inner}$ and $\theta_{\rm outer}$, such that spaxels contribute only if
\begin{equation}
\theta_{\rm inner} \le \theta_{\rm outflow} \le \theta_{\rm outer}
\quad \text{or} \quad
\pi - \theta_{\rm outer} \le \theta_{\rm outflow} \le \pi - \theta_{\rm inner},
\end{equation}

for $\theta_{\rm outflow}$ in radians. The radial velocity field follows an empirical prescription, with a linear rise to a turnover radius $R_{\rm outflow}$ followed by a linear decline, for which we adopt the following empirical form:

\begin{equation}
v_{\rm rad}(R^*) =
\begin{cases}
v_{\rm outflow} \frac{R^*}{R_{\rm outflow}}, & R^* \le R_{\rm outflow}, \\
v_{\rm outflow}\left(2 - \frac{R^*}{R_{\rm outflow}}\right), & R_{\rm outflow} < R^* \le 2R_{\rm outflow} \\
0, & R^* > 2R_{\rm outflow}.
\end{cases}
\end{equation}

The observed velocity is then approximated by scaling the radial velocity according to the angular separation from the outflow axis,

\begin{equation}
\label{eq:outflow_model}
v_{\rm outflow-model}(x,y) =
v_{\rm sys} + v_{\rm rad}(R^*)\cos\theta_{\rm outflow},
\end{equation}

where emission outside the bicone is set to zero. For the outflow, we adopt $i_{\rm outflow} = 48^\circ$ from \citep{shimizu2019multiphase} and find $\phi_{\rm outflow} = 110^\circ$ east from north. We fix these values and fit for the maximum outflow velocity, $v_{\rm outflow}$, turnover radius, $R_{\rm outflow}$, and inner and outer opening angles, $\theta_{\rm inner}$ and $\theta_{\rm outer}$. We consistently find $\theta_{\rm inner} > 0$, supporting a hollow rather than filled ionisation cone.

We also compared our adopted bicone geometry with that derived using \texttt{MOKA3D} \citep{marconcini2023moka3d} (v0.1), with full details provided in the supplementary material. We fit the bicone independently to the high-IP [O IV], [Ne V], and [Ne VI] lines, which most directly trace the outflow and provide the largest high S/N spatial coverage (figure~S2 of the supplementary material). The three lines yield consistent geometries, with mean blue- and red-shifted bicone inclinations of $\beta_{\rm blue} = 83^\circ$ and $\beta_{\rm red} = 98^\circ$, respectively, both placing the bicone close to the plane of the sky. These results are also consistent with the \texttt{MOKA3D} results of \citet{marconcini2025fast}. However, the \texttt{MOKA3D} inclinations, while consistent between our mid-IR analysis and the optical analysis of \citet{marconcini2025fast}, differ from the $\sim48^\circ$ inclination found by \citet{durre2019agn} and \citet{shimizu2019multiphase}, who instead constrained the inclination using 2D projected velocity fields.

We therefore retain the outflow inclination of $\sim48^\circ$ from \citet{durre2019agn} and \citet{shimizu2019multiphase}, as our own optimisation of the projected velocity field (equation~\ref{eq:outflow_model}) across multiple ionic lines also favours an inclination closer to $\sim48^\circ$. This choice is not in conflict with the \texttt{MOKA3D} fits: \texttt{MOKA3D} assumes a constant intrinsic radial velocity, with the range of projected velocities arising from gas at different inclinations within the bicone (the axis inclination $\pm$ the half-opening angle), reaching a maximum of $\sim1180$~km~s$^{-1}$ for gas closest to our line of sight. Applying our kinematic fitting to the \texttt{MOKA3D} model cube reproduces the observed projected velocities well. We emphasise that a detailed comparison of \texttt{MOKA3D} and other outflow or disc modelling approaches is beyond the scope of this study. Our decomposition method can separate the relative emission contributions of the different components for any physically reasonable outflow or disc model, with the appropriate model choice depending on the system being studied.

\subsubsection{MCMC analysis}
\label{sec:MCMC}

We obtain posterior distributions using the affine-invariant MCMC sampler \texttt{emcee} \citep{emcee2013}. Uniform priors are adopted over physically motivated ranges, and parameter estimates are taken from the marginalised posterior distributions (see supplementary material). The resulting best fit parameters are used to construct model velocity fields and residual maps for comparison with the observed IFS kinematics, which we show for the [S III] line as examples in figure~S4 of the supplementary material.

The emission lines listed in Table~\ref{tab:ion_lines} trace each kinematic component to varying degrees. The [Fe II], [Ar II], and [Ne II] lines are dominated by disc emission across most spaxels, whereas [O IV], [Ar VI], and [Mg IV] predominantly trace the outflow. In addition, the [Ne V] and [Ne VI] lines exclusively trace the outflow kinematics (see figures~S1 and S2). These lines therefore provide valuable constraints on the outflow velocity field model, but show no evidence of tracing the star-forming ring in NGC~5728. We therefore fit the former group of three lines using only the disc velocity field model (Section~\ref{sec:disc_model}), and the latter group of five lines using only the outflow model (Section~\ref{sec:bicone_model}). These single component fits provide good overall descriptions, except in spaxels where the other component is non-negligible. This yields estimates of $v_{\rm disc}$ and $R_{\rm disc}$ from the disc dominated lines, and $v_{\rm outflow}$, $R_{\rm outflow}$, $\theta_{\rm inner}$, and $\theta_{\rm outer}$ from the outflow dominated lines, which are consistent within uncertainties across lines within each set of lines. We use these results to inform the priors on the relevant parameters before performing MCMC optimisation across all ten remaining emission lines (now excluding [Ne V] and [Ne VI] from subsequent analysis as they show no disc component) using a combined disc plus outflow velocity field model. The final velocity in each spaxel is expressed as:

\begin{equation}
v(x, y) = \alpha v_{\rm disc-model}(x, y) + (1-\alpha) v_{\rm outflow-model}(x, y), 
\end{equation} 

where $0 \leq \alpha \leq 1$ is a parameter that quantifies the relative disc contribution to the velocity. Within the MCMC framework, $\alpha$ is held constant across all spaxels, while allowed to vary between lines to assess their relative sensitivity to the disc and outflow kinematics.

Running the two-component MCMC fits across the ten emission lines yields parameters that are generally consistent between lines, with the exception of $R_{\rm outflow}$ and $\alpha$, which vary systematically with line IP. For the remaining parameters, we obtain average values of $v_{\rm disc} = 400 \pm 50$ km s$^{-1}$, $R_{\rm disc} = 290 \pm 30$ pc, $v_{\rm outflow} = 700 \pm 90$ km s$^{-1}$, $\theta_{\rm inner} = 21 \pm 5^\circ$, and $\theta_{\rm outer} = 80 \pm 5^\circ$.

For the two parameters that vary significantly with IP, we find approximately linear trends: $R_{\rm outflow} = (3.6 \times {\rm IP}) + 535$ pc\footnote{This implies that in these mid-IR observations, the outflow extends out to $\sim 1000$–$1700$ pc from the AGN; aligning with what previous studies find at other wavelengths (e.g., \citealt{shimizu2019multiphase}).}, for IP in eV, and $\alpha = 0.81 - (0.01 \times {\rm IP})$ for IP $\leq 81$, with $\alpha = 0$ for IP $> 81$. This piecewise behaviour in $\alpha$ reflects our highest IP capable of tracing the disc, [Mg IV] (IP = 80.1 eV), which is found to trace almost exclusively the outflow, motivating $\alpha \simeq 0$ beyond this IP.

The increase of $R_{\rm outflow}$ with IP does not necessarily imply that higher ionisation gas extends to larger physical distances from the AGN. Instead, it likely reflects the decreasing contribution of disc emission with increasing IP, allowing the outflow kinematics to be traced over a larger projected extent. At low IP, emission from the circumnuclear ring dominates at these radii, obscuring the outflow signature, whereas high-IP lines are increasingly dominated by AGN-ionised gas and therefore trace the outflow more clearly to larger distances. Moreover, the decrease of $\alpha$ with IP reflects the transition from low-energy photons ($\lesssim 20$ eV) associated with star formation to high-energy photons ($\gtrsim 55$ eV) produced by AGN-driven processes, naturally explaining the observed trend.

Overall, this MCMC optimisation enables us to disentangle the disc and outflow contributions and construct separate velocity maps for each component in NGC~5728. The resulting combined velocity fields as a function of IP are presented in figure~S5 of the supplementary material. These are consistent with the observed velocity maps shown in figure~S2, which demonstrate the gradual transition from a velocity field dominated by the disc rotation at low IP to one tracing exclusively the outflow at high IP.

\subsection{Component flux fractions}
\label{sec:flux_fractions}

While the parameter $\alpha$ is useful for characterising line-averaged behaviour, it does not reproduce the line profile within individual spaxels. Indeed, allowing $\alpha$ to vary freely per spaxel drives the model to reproduce the observed velocity exactly wherever it lies between $v_{\rm disc-model}$ and $v_{\rm outflow-model}$, a condition that is typically satisfied. Although this yields exceptional fits to the observed velocity maps, it results in a degenerate solution with no physical predictive power, effectively constituting overfitting.

Instead, we use constrained double Gaussian fits to the line profile. The Gaussian centroids are fixed to the velocities predicted by the disc and outflow models in that spaxel, while the dispersions, $\sigma_{\rm disc}$ and $\sigma_{\rm outflow}$, are fixed to the median values for each component and line across all spaxels\footnote{A spatially varying velocity dispersion could also be adopted if available.}. As shown by \citet{bellocchi2019uncertainties} and \citet{veenema2026kinematics}, tightly constraining the velocity dispersion mitigates many degeneracies in double-Gaussian decomposition of emission line data; here we extend this approach by additionally fixing the centroid velocities to the predicted disc/outflow models. We then fit only the amplitudes, $A_{\rm disc}$ and $A_{\rm outflow}$, for each component in every spaxel and emission line. We define the ``disc fraction'', $f_{\rm disc}$, of a given spaxel and emission line as:

\begin{equation}
f_{\rm disc} = \frac{A_{\rm disc} \sigma_{\rm disc}}{A_{\rm disc} \sigma_{\rm disc} + A_{\rm outflow} \sigma_{\rm outflow}}.
\end{equation} 

This is equal to the fractional flux contributed by the disc component, with the outflow fraction defined as $f_{\rm outflow} = 1 - f_{\rm disc}$. This construction decomposes the flux in each component and enables the production of physically meaningful $f_{\rm disc}$ and $f_{\rm outflow}$ maps for each emission line. Our method uses kinematics to separate components and then imposes flux decomposition to assign emission to each, thereby reducing degeneracies associated with traditional decomposition approaches.

We apply this procedure to all ten remaining emission lines in Table~\ref{tab:ion_lines} for NGC~5728, with the resulting $f_{\rm disc}$ maps presented in figure~S6 of the supplementary material. The maps are broadly consistent across lines (with a median line-to-line scatter of only 0.17 in the inferred disc fraction), showing high $f_{\rm disc}$ within and around the circumnuclear star-forming ring, and low $f_{\rm disc}$ (high $f_{\rm outflow}$) at the AGN and outflow.

\subsection{Stacked approach}
\label{sec:stacked_approach}

Since all remaining emission lines trace both components to varying degrees, we construct a weighted average of $f_{\rm disc}$ across all spaxels. We assign higher weights to lines that trace both components comparably, as these provide more robust constraints. To do this, all $f_{\rm disc}$ maps are first reprojected onto a common spatial grid using \texttt{reproject\_interp} \citep{robitaille2020reproject}, after which a weighted sum in $f_{\rm disc}$ in every spaxel is computed. The weights are defined by each line's mean $f_{\rm disc}$ across the FOV, with maximum weight assigned at $f_{\rm disc}=0.5$ and decreasing linearly to zero at $f_{\rm disc}=0$ or $1$. The resulting average $f_{\rm disc}$ values and corresponding weights are listed in Table~\ref{tab:ion_lines}. We restrict the weighted average to spaxels where at least two emission lines are available, masking out all others to minimise potential biasses from any individual line on the final result.

\begin{figure}
   \centering
   \includegraphics[width=\columnwidth]{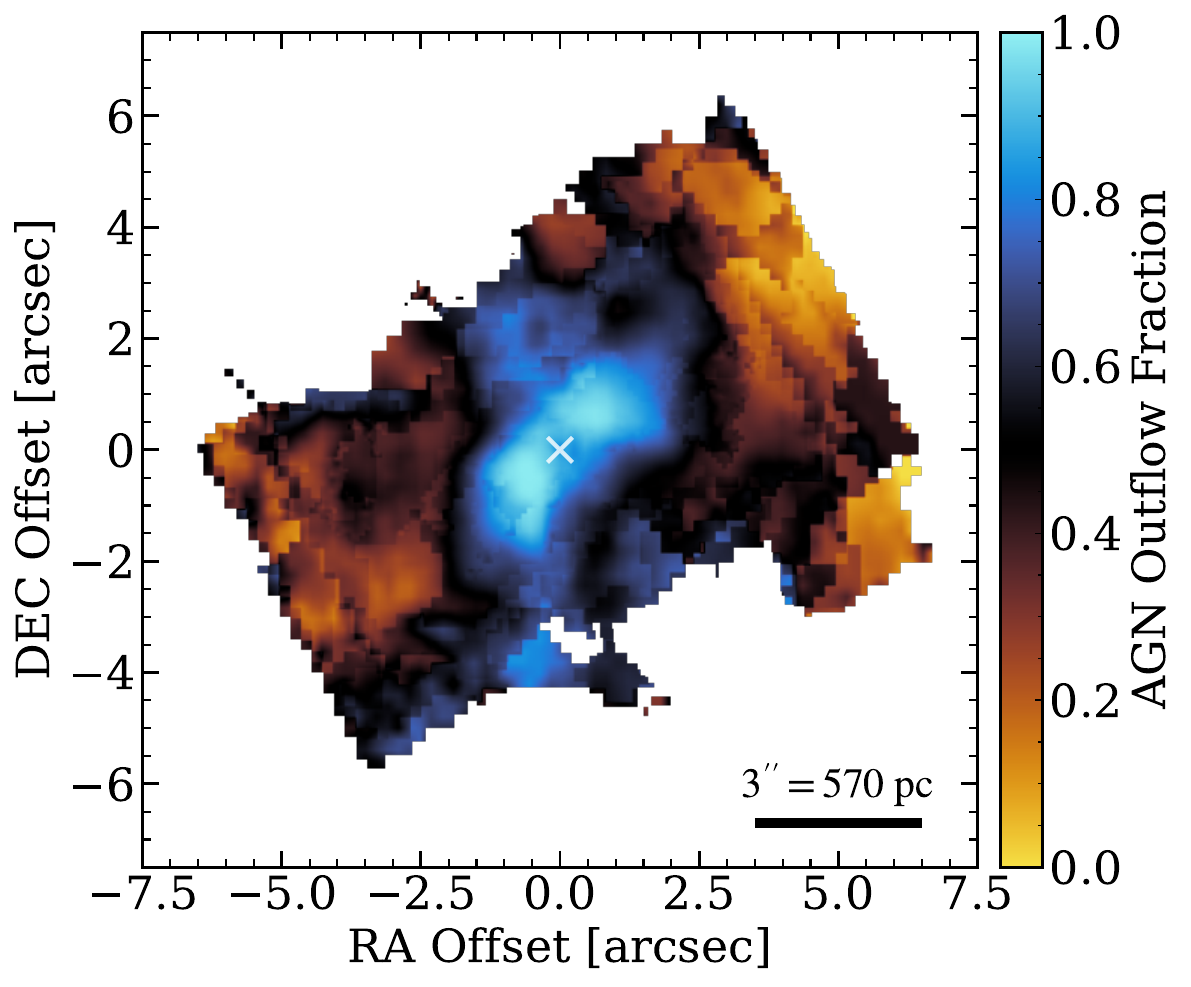}
   \caption{Outflow flux fraction map, average $f_{\rm outflow}$, of NGC~5728 derived using our multi-emission line stacked kinematics framework. Regions with high $f_{\rm outflow}$ have flux emission dominated by the AGN-driven outflow, while regions with low $f_{\rm outflow}$ are dominated by the circumnuclear star-forming disc. North is up and east is to the left. The cross denotes the AGN position.}
    \label{fig:outflow_frac}%
\end{figure}

This procedure yields mean $f_{\rm disc}$ and $f_{\rm outflow}$ maps that combine information from multiple tracers, providing a spatially resolved view of both components. We show the resulting $f_{\rm outflow}$ map for NGC~5728 in Fig.~(\ref{fig:outflow_frac}). The robustness of this stacked map is verified by repeating the $f_{\rm disc}$ or $f_{\rm outflow}$ averaging with alternative weighting schemes, including equal weights, Gaussian-weighted schemes centred on $f_{\rm disc}=0.5$, use of the median instead of the mean, and by selectively including or excluding subsets of emission lines. In all cases, the resulting $f_{\rm outflow}$ maps are consistent with that shown in Fig.~(\ref{fig:outflow_frac}), demonstrating that the solution is stable and not sensitive to the specific weighting or individual emission line choices.

\section{Discussion}
\label{sec:Discussion}

\subsection{Application to NGC~5728}
\label{sec: application_ngc5728}

Fig.~(\ref{fig:outflow_frac}) shows the spatial variation of the average $f_{\rm outflow}$ across the nuclear and circumnuclear regions of NGC~5728. Regions with $f_{\rm outflow} \gtrsim 0.8$ strongly trace the AGN and its known biconical outflow, while values decrease towards the known circumnuclear star-forming ring. Within the ring, $f_{\rm outflow} \lesssim 0.2$, demonstrating that the method effectively separates the two components and spatially disentangles their emission. The outflow morphology, position angle, and location of the outflow and ring are consistent with previous studies (e.g., \citealt{wilson1993, shimizu2019multiphase, durre2019agn, HermosaMunoz2026}), supporting the reliability of our results.

Furthermore, we find lower $f_{\rm outflow}$ in the western portion of the ring compared to the eastern side, indicating stronger AGN contribution to the emission on the eastern side. The outflow emission is also most prominent symmetrically about the nucleus along the major bicone axis out to $\sim 500$ pc on either side, beyond which the emission becomes increasingly mixed with star-formation dominated regions. A region north of the AGN, approximately perpendicular to the main bicone axis, shows enhanced $f_{\rm outflow} \gtrsim 0.7$, indicating a strong AGN contribution. This is consistent with \citet{GarciaBernete2024PAH}, who identify potential AGN-excited emission in this region, and \citet{davies2024gatos}, who find enhanced ionised gas velocity dispersion, likely associated with AGN activity. One possible explanation is a lateral outflow driven by interaction between the main AGN outflow and the surrounding ISM, analogous to the jet driven lateral outflow proposed by \citet{audibert2023jet}, where the jet compresses and accelerates gas perpendicular to its axis. Overall, this method robustly separates the AGN outflow and star-forming ring, while also identifying intermediate regions where both components contribute significantly and almost equally to the observed emission.

\subsection{Further applications}
\label{sec:further_applications}

Beyond its application to these JWST observations of NGC~5728, this method has several additional potential applications, which we discuss in this subsection.

\subsubsection{Identifying additional components}

In Section~\ref{sec:flux_fractions} we introduced flux decomposition by fitting the component amplitudes in the constrained double-Gaussian model, with centroid velocities and velocity dispersions fixed. An additional unmodelled component would produce systematic residual flux, which could be mapped per spaxel to diagnose its spatial distribution and motivate extending the model. For NGC~5728, however, we find no evidence for additional components: the constrained fits reproduce the line profiles well, with residual maps showing only low-level, spatially incoherent noise across all emission lines in the FOV. This method could be extended to more components provided their velocity fields and dispersions are sufficiently constrained and distinguishable, and multiple tracers are available for each component.

\subsubsection{Comparing spectra of star-forming to AGN dominated regions}

The $f_{\rm outflow}$ maps can be used to identify regions dominated by individual components for spectral comparison. For example, selecting spaxels with average $f_{\rm outflow} > 0.8$ and $f_{\rm outflow} < 0.2$ provides AGN-dominated and star-forming spectra, respectively, enabling the construction of representative ``pure'' templates across any wavelength range. Applying this approach to a galaxy sample would allow the creation of component specific composite spectra and the development of new data-driven diagnostics for separating stellar and AGN activity. We demonstrate the broader applicability of our technique to additional targets observed with JWST in future work (Veenema et al., in prep.).

Our method can quantify the relative contributions of star formation and AGN-driven processes to dust heating in the circumnuclear region. The 21 $\mu$m continuum is a commonly used tracer of warm dust emission \citep{lopez2025gatos, chown2025}. Multiplying the averaged $f_{\rm outflow}$ map (Fig.~\ref{fig:outflow_frac}) by the 21 $\mu$m continuum map (figure~S7 of the supplementary material) provides a proxy for the outflow contribution to the warm dust emission, with the complementary map giving the corresponding star-forming contribution. This interpretation assumes that the spatial fractions derived from the ionised gas provide a reasonable proxy for the relative contributions of the outflow and disc to the warm dust emission. The disc contribution is predominantly confined to the ring, but also shows enhanced emission along a north-east to south-west axis through the AGN, approximately coincident with the nuclear stellar bar identified by \citet{schommer1988ionized} and \citet{prada1999counterrotating}, and shown in figure~5 of \citet{shimizu2019multiphase}. Our method therefore identifies the nuclear stellar bar in the 21 $\mu$m emission and associates it with the star-forming component. Within the JWST field of view, which covers most but not all of the circumnuclear disc, at least 62\% of the 21 $\mu$m emission is attributed to the AGN outflow under this assumption, indicating that AGN heating dominates the warm dust emission over the observed region. This is consistent with \citet{haidar2026gatos}, who identify excess dust heating along the ionisation cones of NGC~5728 and suggest shocks associated with the outflow as an additional heating mechanism. Our decomposition therefore provides a proxy for disentangling the AGN and star-forming contributions to the observed nuclear dust emission.

\subsubsection{Potential limitations}

While NGC~5728 provides a useful demonstration target due to its well defined circumnuclear ring and biconical outflow, we will test the generality of this method across a wider range of targets in future work. Potential limitations include degeneracies between the velocity and velocity dispersion of each component simultaneously causing erroneous decomposition, non-biconical outflow geometries requiring a different model velocity field, systems with more than two kinematic components requiring an extension of the method, and targets with fewer emission line tracers for all components. In the latter case, the reduced information available for stacking could bias the resulting component fraction maps towards the properties of the available tracers. Our method requires sufficient spatial resolution to resolve the relevant components across enough spaxels to constrain their kinematics, and sufficient spectral resolution to distinguish spatially overlapping components through their velocity and velocity dispersion profiles.

\section{Conclusions}
\label{sec:conclusions}

We present a general framework for decomposing spatially resolved emission in the circumnuclear regions of AGN to isolate and map distinct physical components on a spaxel-by-spaxel basis. This method models or constrains the velocity field and velocity dispersion of each component, and then optimises their amplitudes within a constrained double-Gaussian formalism that fits the emission line profile for each spaxel. Applied across multiple emission lines, the resulting component flux fractions are combined via a weighted average to yield a robust quantification of each component's spatial distribution and relative contribution to the observed flux emission.

We demonstrate the power of this method using twelve mid-IR ionic lines in JWST IFS observations of NGC~5728 from the Galaxy Activity, Torus, and Outflow Survey (GATOS). The emission can be well reproduced by two components: a circumnuclear star-forming ring and an AGN-driven biconical outflow. The resulting flux fraction maps provide high fidelity maps of the morphology of these components, enabling their properties to be studied independently. We use this to calculate that $\sim 62 \%$ of the 21 $\mu$m luminosity is attributable to the AGN within the JWST FOV, assuming that the warm dust also follows our inferred outflow fraction.

More broadly, this framework is fully general and can be applied to any IFS dataset with multiple source components across any wavelength regime, provided suitable tracers exist. This is particularly relevant for studies of active galaxies, where star-formation and AGN driven processes frequently overlap in both spectral and spatial domains, making robust component separation essential.

\section*{Acknowledgements}

OV is supported by a Science and Technology Facilities Council (STFC) studentship No. ST/Y509474/1. DR and NT acknowledge the support from a Leverhulme Trust Research Project Grant. IGB is supported by the Programa Atracci\'on de Talento Investigador ``C\'esar Nombela'' via grant 2023-T1/TEC-29030 funded by the Community of Madrid, and acknowledges support from the research project PID2024-159902NA-I00 funded by the Spanish Ministry of Science and Innovation/State Agency of Research (MCIN/AEI/10.13039/501100011033) and FSE+. SGB and FE acknowledge support from the Spanish grant PID2022-138560NB-I00, funded by MCIN/AEI/10.13039/501100011033/FEDER, EU. 
MPS acknowledges support under grants RYC2021-033094-I, CNS2023-145506, and PID2023-146667NB-I00 funded by MCIN/AEI/10.13039/501100011033 and the European Union NextGenerationEU/PRTR.
AJB acknowledges funding from the ``FirstGalaxies'' Advanced Grant from the European Research Council (ERC) under the European Union's Horizon 2020 research and innovation programme (Grant agreement No.\ 789056).
EB acknowledges support from the Spanish grants PID2022-138621NB-I00 and PID2021-123417OB-I00, funded by MCIN/AEI/10.13039/501100011033/FEDER, EU.
CRA and AA acknowledge support from the Agencia Estatal de Investigaci\'on of the Ministerio de Ciencia, Innovaci\'on y Universidades (MCIU/AEI) under the grant ``Tracking active galactic nuclei feedback from parsec to kiloparsec scales'', with reference PID2022--141105NB--I00 and the European Regional Development Fund (ERDF).
AA acknowledges support from the European Union (WIDERA ExGal-Twin, GA~101158446).
SFH acknowledges support through UK Research and Innovation (UKRI) under the UK government’s Horizon Europe Funding Guarantee (EP/Z533920/1, selected in the 2023 ERC Advanced Grant round) and an STFC Small Award (ST/Y001656/1).
AAH acknowledges support from grant PID2021-124665NB-I00 funded by the Spanish Ministry of Science and Innovation and the State Agency of Research MCIN/AEI/10.13039/501100011033 and ERDF (``A way of making Europe'').
OG-M acknowledges the support received by the UNAM DGAPA-PAPIIT project IN109123 and SEHCITI Ciencia de Frontera project CF-2023-G100. 
MS was supported by the Ministry of Science, Technological Development and Innovation of the Republic of Serbia (MSTDIRS) through contract no. 451-03-33/2026-03/200002 with the Astronomical Observatory (Belgrade).
RAR acknowledges the support from the Conselho Nacional de Desenvolvimento Cient'ifico e Tecnol'ogico (CNPq; Projects 303450/2022-3 and 403398/2023-1), the Coordena\c{c}~ao de Aperfei\c{c}oamento de Pessoal de N'ivel Superior (CAPES; Project 88887.894973/2023-00), and Funda\c{c}~ao de Amparo `a Pesquisa do Estado do Rio Grande do Sul (FAPERGS; Project 25/2551-0002765-9).

\section*{Data Availability}

These data were obtained from the Mikulski Archive for Space Telescopes (MAST) at the Space Telescope Science Institute. These observations are associated with programs \#1670 for MIRI/MRS and \#5017 for NIRSpec IFU.



\bibliographystyle{mnras}
\bibliography{references}

@article{kawara1988forbidden,
  title={Forbidden Fe II 1.644 micron emission line in Seyfert and starburst galaxies},
  author={Kawara, Kimiaka and Nishida, Minoru and Taniguchi, Yoshiaki},
  journal={Astrophysical Journal, Part 2-Letters (ISSN 0004-637X), vol. 328, May 15, 1988, p. L41-L44.},
  volume={328},
  pages={L41--L44},
  year={1988}
}

@article{pereira2024extended,
  title={Extended high-ionization [Mg IV] emission tracing widespread shocks in starbursts seen by JWST/NIRSpec},
  author={Pereira-Santaella, Miguel and Garc{\'\i}a-Bernete, Ismael and Gonz{\'a}lez-Alfonso, Eduardo and Alonso-Herrero, Almudena and Colina, Luis and Garc{\'\i}a-Burillo, Santiago and Rigopoulou, Dimitra and Arribas, Santiago and Perna, Michele},
  journal={Astronomy \& Astrophysics},
  volume={685},
  pages={L13},
  year={2024},
  publisher={EDP Sciences}
}

@article{davies2024gatos,
  title={GATOS: missing molecular gas in the outflow of NGC 5728 revealed by JWST},
  author={Davies, R and Shimizu, T and Pereira-Santaella, Miguel and Alonso-Herrero, A and Audibert, A and Bellocchi, E and Boorman, P and Campbell, S and Cao, Y and Combes, F and others},
  journal={Astronomy \& Astrophysics},
  volume={689},
  pages={A263},
  year={2024},
  publisher={EDP Sciences}
}

@article{munoz2024biconical,
  title={A biconical ionised gas outflow and evidence of positive feedback in NGC 7172 uncovered by MIRI/JWST},
  author={Hermosa Mu{\~n}oz, L and Alonso-Herrero, A and Pereira-Santaella, Miguel and Garc{\'\i}a-Bernete, I and Garc{\'\i}a-Burillo, S and Garc{\'\i}a-Lorenzo, B and Davies, R and Shimizu, T and Esparza-Arredondo, D and Hicks, EKS and others},
  journal={Astronomy \& Astrophysics},
  volume={690},
  pages={A350},
  year={2024},
  publisher={EDP Sciences}
}

@article{kewley2019understanding,
  title={Understanding galaxy evolution through emission lines},
  author={Kewley, Lisa J and Nicholls, David C and Sutherland, Ralph S},
  journal={Annual Review of Astronomy and Astrophysics},
  volume={57},
  number={1},
  pages={511--570},
  year={2019},
  publisher={Annual Reviews}
}

@article{Jakobsen_2022,
   title={The Near-Infrared Spectrograph (NIRSpec) on theJames WebbSpace Telescope: I. Overview of the instrument and its capabilities},
   volume={661},
   ISSN={1432-0746},
   url={http://dx.doi.org/10.1051/0004-6361/202142663},
   DOI={10.1051/0004-6361/202142663},
   journal={Astronomy &amp; Astrophysics},
   publisher={EDP Sciences},
   author={Jakobsen, P. and Ferruit, P. and Alves de Oliveira, C. and Arribas, S. and Bagnasco, G. and Barho, R. and Beck, T. L. and Birkmann, S. and Böker, T. and Bunker, A. J. and Charlot, S. and de Jong, P. and de Marchi, G. and Ehrenwinkler, R. and Falcolini, M. and Fels, R. and Franx, M. and Franz, D. and Funke, M. and Giardino, G. and Gnata, X. and Holota, W. and Honnen, K. and Jensen, P. L. and Jentsch, M. and Johnson, T. and Jollet, D. and Karl, H. and Kling, G. and Köhler, J. and Kolm, M.-G. and Kumari, N. and Lander, M. E. and Lemke, R. and López-Caniego, M. and Lützgendorf, N. and Maiolino, R. and Manjavacas, E. and Marston, A. and Maschmann, M. and Maurer, R. and Messerschmidt, B. and Moseley, S. H. and Mosner, P. and Mott, D. B. and Muzerolle, J. and Pirzkal, N. and Pittet, J.-F. and Plitzke, A. and Posselt, W. and Rapp, B. and Rauscher, B. J. and Rawle, T. and Rix, H.-W. and Rödel, A. and Rumler, P. and Sabbi, E. and Salvignol, J.-C. and Schmid, T. and Sirianni, M. and Smith, C. and Strada, P. and te Plate, M. and Valenti, J. and Wettemann, T. and Wiehe, T. and Wiesmayer, M. and Willott, C. J. and Wright, R. and Zeidler, P. and Zincke, C.},
   year={2022},
   month=may, pages={A80} }

@article{wells2015mid,
  title={The mid-infrared instrument for the james webb space telescope, vi: The medium resolution spectrometer},
  author={Wells, Martyn and Pel, J-W and Glasse, Alistair and Wright, GS and Aitink-Kroes, Gabby and Azzollini, Ruym{\'a}n and Beard, Steven and Brandl, BR and Gallie, Angus and Geers, VC and others},
  journal={Publications of the Astronomical Society of the Pacific},
  volume={127},
  number={953},
  pages={646},
  year={2015},
  publisher={IOP Publishing}
}

@article{argyriou2023jwst,
  title={JWST MIRI flight performance: the medium-resolution spectrometer},
  author={Argyriou, Ioannis and Glasse, Alistair and Law, David R and Labiano, Alvaro and {\'A}lvarez-M{\'a}rquez, Javier and Patapis, Polychronis and Kavanagh, Patrick J and Gasman, Danny and Mueller, Michael and Larson, Kirsten and others},
  journal={Astronomy \& Astrophysics},
  volume={675},
  pages={A111},
  year={2023},
  publisher={EDP Sciences}
}

@article{thornley2000massive,
  title={Massive Star Formation and Evolution in Starburst Galaxies: Mid-infrared Spectroscopy with the ISO Short WavelengthSpectrometer},
  author={Thornley, Michele D and Schreiber, Natascha M F{\"o}rster and Lutz, Dieter and Genzel, Reinhard and Spoon, Henrik WW and Kunze, Dietmar and Sternberg, Amiel},
  journal={The Astrophysical Journal},
  volume={539},
  number={2},
  pages={641},
  year={2000},
  publisher={IOP Publishing}
}

@article{cappellari2023full,
  title={Full spectrum fitting with photometry in PPXF: stellar population versus dynamical masses, non-parametric star formation history and metallicity for 3200 LEGA-C galaxies at redshift z≈ 0.8},
  author={Cappellari, Michele},
  journal={Monthly Notices of the Royal Astronomical Society},
  volume={526},
  number={3},
  pages={3273--3300},
  year={2023},
  publisher={Oxford University Press}
}

@article{robitaille2020reproject,
  title={reproject: Python-based astronomical image reprojection},
  author={Robitaille, Thomas and Deil, Christoph and Ginsburg, Adam},
  journal={Astrophysics Source Code Library},
  pages={ascl--2011},
  year={2020}
}

@article{garcia2024galaxy,
  title={The Galaxy Activity, Torus, and Outflow Survey (GATOS)-III. Revealing the inner icy structure in local active galactic nuclei},
  author={Garc{\'\i}a-Bernete, I and Alonso-Herrero, A and Rigopoulou, D and Pereira-Santaella, Miguel and Shimizu, T and Davies, R and Donnan, FR and Roche, PF and Gonz{\'a}lez-Mart{\'\i}n, O and Almeida, C Ramos and others},
  journal={Astronomy \& Astrophysics},
  volume={681},
  pages={L7},
  year={2024},
  publisher={EDP Sciences}
}

@article{vivian2022goals,
  title={GOALS-JWST: Resolving the Circumnuclear Gas Dynamics in NGC 7469 in the Mid-infrared},
  author={U, Vivian and Lai, Thomas and Bianchin, Marina and Remigio, Raymond P and Armus, Lee and Larson, Kirsten L and D{\'\i}az-Santos, Tanio and Evans, Aaron and Stierwalt, Sabrina and Law, David R and others},
  journal={The Astrophysical Journal Letters},
  volume={940},
  number={1},
  pages={L5},
  year={2022},
  publisher={IOP Publishing}
}

@article{garcia2021galaxy,
  title={The Galaxy Activity, Torus, and Outflow Survey (GATOS)-I. ALMA images of dusty molecular tori in Seyfert galaxies},
  author={Garc{\'\i}a-Burillo, S and Alonso-Herrero, A and Almeida, C Ramos and Gonz{\'a}lez-Mart{\'\i}n, O and Combes, F and Usero, A and H{\"o}nig, S and Querejeta, M and Hicks, EKS and Hunt, Leslie Kipp and others},
  journal={Astronomy \& Astrophysics},
  volume={652},
  pages={A98},
  year={2021},
  publisher={EDP Sciences}
}

@article{alonso2021galaxy,
  title={The Galaxy Activity, Torus, and Outflow Survey (GATOS)-II. Torus and polar dust emission in nearby Seyfert galaxies},
  author={Alonso-Herrero, A and Garc{\'\i}a-Burillo, S and H{\"o}nig, SF and Garc{\'\i}a-Bernete, I and Almeida, C Ramos and Gonz{\'a}lez-Mart{\'\i}n, O and L{\'o}pez-Rodr{\'\i}guez, E and Boorman, PG and Bunker, AJ and Burtscher, L and others},
  journal={Astronomy \& Astrophysics},
  volume={652},
  pages={A99},
  year={2021},
  publisher={EDP Sciences}
}

@article{boker2022near,
  title={The near-infrared spectrograph (nirspec) on the james webb space telescope-iii. integral-field spectroscopy},
  author={B{\"o}ker, Torsten and Arribas, S and L{\"u}tzgendorf, N and de Oliveira, C Alves and Beck, TL and Birkmann, S and Bunker, AJ and Charlot, S and De Marchi, G and Ferruit, P and others},
  journal={Astronomy \& Astrophysics},
  volume={661},
  pages={A82},
  year={2022},
  publisher={EDP Sciences}
}

@article{pereira2010mid,
  title={The mid-infrared high-ionization lines from active galactic nuclei and star-forming galaxies},
  author={Pereira-Santaella, Miguel and Diamond-Stanic, Aleksandar M and Alonso-Herrero, Almudena and Rieke, George H},
  journal={The Astrophysical Journal},
  volume={725},
  number={2},
  pages={2270},
  year={2010},
  publisher={IOP Publishing}
}

@article{bianchin2024goals,
  title={GOALS-JWST: Gas Dynamics and Excitation in NGC 7469 Revealed by NIRSpec},
  author={Bianchin, Marina and Vivian, U and Song, Yiqing and Lai, Thomas S-Y and Remigio, Raymond P and Barcos-Mu{\~n}oz, Loreto and D{\'\i}az-Santos, Tanio and Armus, Lee and Inami, Hanae and Larson, Kirsten L and others},
  journal={The Astrophysical Journal},
  volume={965},
  number={2},
  pages={103},
  year={2024},
  publisher={IOP Publishing}
}

@article{colina2015understanding,
  title={Understanding the two-dimensional ionization structure in luminous infrared galaxies-A near-IR integral field spectroscopy perspective},
  author={Colina, Luis and L{\'o}pez, Javier Piqueras and Arribas, Santiago and Riffel, Rog{\'e}rio and Riffel, Rogemar A and Rodriguez-Ardila, Alberto and Pastoriza, Miriani and Storchi-Bergmann, Thaisa and Alonso-Herrero, Almudena and Sales, Dinalva},
  journal={Astronomy \& Astrophysics},
  volume={578},
  pages={A48},
  year={2015},
  publisher={EDP Sciences}
}

@article{almeida2025jwst,
  title={JWST MIRI reveals the diversity of nuclear mid-infrared spectra of nearby type 2 quasars},
  author={Ramos Almeida, C and Garc{\'\i}a-Bernete, I and Pereira-Santaella, M and Speranza, G and Maiolino, R and Ji, X and Audibert, A and Cezar, PH and Acosta-Pulido, JA and Alonso-Herrero, A and others},
  journal={Astronomy \& Astrophysics},
  volume={698},
  pages={A194},
  year={2025},
  publisher={EDP Sciences}
}

@article{schommer1988ionized,
  title={Ionized gas and radio emission in the barred Seyfert galaxy NGC 5728},
  author={Schommer, Robert A and Caldwell, Nelson and Wilson, AS and Baldwin, JA and Phillips, MM and Williams, TB and Turtle, AJ},
  journal={Astrophysical Journal, Part 1 (ISSN 0004-637X), vol. 324, Jan. 1, 1988, p. 154-171.},
  volume={324},
  pages={154--171},
  year={1988}
}

@article{shimizu2019multiphase,
  title={The multiphase gas structure and kinematics in the circumnuclear region of NGC 5728},
  author={Shimizu, T Taro and Davies, RI and Lutz, D and Burtscher, L and Lin, M and Baron, D and Davies, RL and Genzel, R and Hicks, EKS and Koss, M and others},
  journal={Monthly Notices of the Royal Astronomical Society},
  volume={490},
  number={4},
  pages={5860--5887},
  year={2019},
  publisher={Oxford University Press}
}

@article{durre2019agn,
  title={The AGN ionization cones of NGC 5728. II. Kinematics},
  author={Durr{\'e}, Mark and Mould, Jeremy},
  journal={The Astrophysical Journal},
  volume={870},
  number={1},
  pages={37},
  year={2019},
  publisher={IOP Publishing}
}

@ARTICLE{herrero2025miconic,
       author = {{Alonso-Herrero}, A. and {Hermosa Mu{\~n}oz}, L. and {Labiano}, A. and {Guillard}, P. and {Garc{\'\i}a-Mar{\'\i}n}, M. and {Dicken}, D. and {Garc{\'\i}a-Burillo}, S. and {Pantoni}, L. and {Buiten}, V. and {Colina}, L. and {B{\"o}ker}, T. and {Baes}, M. and {Eckart}, A. and {Evangelista}, L. and {{\"O}stlin}, G. and {Rouan}, D. and {van der Werf}, P. and {Walter}, F. and {Ward}, M.~J. and {Wright}, G. and {G{\"u}del}, M. and {Henning}, Th. and {Lagage}, P.-O.},
        title = "{MICONIC: JWST/MIRI MRS reveals a fast ionized gas outflow in the central region of Centaurus A}",
      journal = {\aap},
         year = 2025,
        month = jul,
       volume = {699},
          eid = {A334},
        pages = {A334},
          doi = {10.1051/0004-6361/202554823},
archivePrefix = {arXiv},
       eprint = {2506.15286},
 primaryClass = {astro-ph.GA},
       adsurl = {https://ui.adsabs.harvard.edu/abs/2025A&A...699A.334A}
}

@article{marconcini2023moka3d,
  title={MOKA3D: An innovative approach to 3D gas kinematic modelling-I. Application to AGN ionised outflows},
  author={Marconcini, C and Marconi, A and Cresci, G and Venturi, G and Ulivi, L and Mannucci, F and Belfiore, F and Tozzi, G and Ginolfi, M and Marasco, A and others},
  journal={Astronomy \& Astrophysics},
  volume={677},
  pages={A58},
  year={2023},
  publisher={EDP Sciences}
}

@article{teodoro20153d,
  title={3D BAROLO: a new 3D algorithm to derive rotation curves of galaxies},
  author={Di Teodoro, EM and Fraternali, Filippo},
  journal={Monthly Notices of the Royal Astronomical Society},
  volume={451},
  number={3},
  pages={3021--3033},
  year={2015},
  publisher={Oxford University Press}
}

@article{ulivi2025jwst,
  title={JWST/NIRSpec insights into the circumnuclear region of Arp 220: A detailed kinematic study},
  author={Ulivi, Lorenzo and Perna, Michele and Lamperti, Isabella and Arribas, Santiago and Cresci, Giovanni and Marconcini, Cosimo and Del Pino, Bruno Rodr{\'\i}guez and B{\"o}ker, Torsten and Bunker, Andrew J and Ceci, Matteo and others},
  journal={Astronomy \& Astrophysics},
  volume={693},
  pages={A36},
  year={2025},
  publisher={EDP Sciences}
}

@article{veenema2025shock,
  title={Shock-driven heating in the circumnuclear star-forming regions of NGC 7582: Insights from JWST NIRSpec and MIRI/MRS spectroscopy},
  author={Veenema, Oscar and Thatte, Niranjan and Rigopoulou, Dimitra and Garc{\'\i}a-Bernete, Ismael and Alonso-Herrero, Almudena and Audibert, Anelise and Bellocchi, Enrica and Bunker, Andrew J and Campbell, Steph and Combes, Francoise and others},
  journal={Monthly Notices of the Royal Astronomical Society},
  volume={544},
  number={4},
  pages={3361--3378},
  year={2025},
  publisher={Oxford University Press}
}

@article{das2005mapping,
  title={Mapping the kinematics of the narrow-line region in the Seyfert galaxy NGC 4151},
  author={Das, Varendra and Crenshaw, DM and Hutchings, JB and Deo, RP and Kraemer, SB and Gull, TR and Kaiser, ME and Nelson, CH and Weistrop, D},
  journal={The Astronomical Journal},
  volume={130},
  number={3},
  pages={945},
  year={2005},
  publisher={IOP Publishing}
}

@misc{kramidateam,
  author = {Kramida, A. and Ralchenko, Y. and Reader, J.},
  title  = {NIST Atomic Spectra Database (version 5.10)},
  year   = {2022},
  howpublished = {National Institute of Standards and Technology, Gaithersburg, MD}
}

@article{zhang2025theoretical,
  title={Theoretical Diagnostics for the Physical Conditions in Active Galactic Nuclei under the View of JWST},
  author={Zhang, Lulu and Davies, Ric I and Packham, Chris and Hicks, Erin KS and Delaney, Daniel E and Pereira-Santaella, Miguel and Mu{\~n}oz, Laura Hermosa and Garc{\'\i}a-Bernete, Ismael and Ricci, Claudio and Rigopoulou, Dimitra and others},
  journal={The Astrophysical Journal Supplement Series},
  volume={280},
  number={2},
  pages={65},
  year={2025},
  publisher={IOP Publishing}
}

@article{zhang2024galaxy,
  title={The Galaxy Activity, Torus, and Outflow Survey (GATOS). IV. Exploring Ionized Gas Outflows in Central Kiloparsec Regions of GATOS Seyferts},
  author={Zhang, Lulu and Packham, Chris and Hicks, Erin KS and Davies, Ric I and Shimizu, Taro T and Alonso-Herrero, Almudena and Mu{\~n}oz, Laura Hermosa and Garc{\'\i}a-Bernete, Ismael and Pereira-Santaella, Miguel and Audibert, Anelise and others},
  journal={The Astrophysical Journal},
  volume={974},
  number={2},
  pages={195},
  year={2024},
  publisher={IOP Publishing}
}

@article{bellocchi2019uncertainties,
  title={Uncertainties in gas kinematics arising from stellar continuum modeling in integral field spectroscopy data: the case of NGC 2906 observed with VLT/MUSE},
  author={Bellocchi, E and Ascasibar, Y and Galbany, L and S{\'a}nchez, SF and Ibarra--Medel, H and Gavil{\'a}n, M and D{\'\i}az, {\'A}},
  journal={Astronomy \& Astrophysics},
  volume={625},
  pages={A83},
  year={2019},
  publisher={EDP Sciences}
}

@article{marconcini2025fast,
  title = {Evidence of the fast acceleration of AGN-driven winds at kiloparsec scales},
  author = {Marconcini, C. and Marconi, A. and Cresci, G. and et al.},
  journal = {Nature Astronomy},
  volume = {9},
  pages = {907--915},
  year = {2025},
  doi = {10.1038/s41550-025-02518-6},
  url = {https://doi.org/10.1038/s41550-025-02518-6}
}

@article{veenema2026kinematics,
  title={Decoupling the AGN outflow and star-forming disc kinematics in the nuclear region of NGC 7582 with JWST NIRSpec and MIRI/MRS},
  author={Veenema, Oscar and Thatte, Niranjan and Rigopoulou, Dimitra and Garc{\'\i}a-Bernete, Ismael and Alonso-Herrero, Almudena and Pereira-Santaella, Miguel and Audibert, Anelise and Bellocchi, Enrica and Bunker, Andrew J and Campbell, Steph and others},
  journal={Monthly Notices of the Royal Astronomical Society},
  volume={548},
  number={4},
  pages={stag785},
  year={2026},
  publisher={Oxford University Press}
}

@ARTICLE{Donnan2026,
       author = {{Donnan}, Fergus R. and {Garc{\'\i}a-Bernete}, Ismael and {Rigopoulou}, Dimitra and {Alonso-Herrero}, Almudena and {Audibert}, Anelise and {Bellocchi}, Enrica and {Bunker}, Andrew and {Campbell}, Steph and {Combes}, Fran{\c{c}}oise and {Davies}, Richard and {D{\'\i}az-Santos}, Tanio and {Fern{\'a}ndez-Ontiveros}, Juan A. and {Gandhi}, Poshak and {Garc{\'\i}a-Burillo}, Santiago and {Gonz{\'a}lez-Mart{\'\i}n}, Omaira and {Hicks}, Erin K.~S. and {Hermosa Mu{\~n}oz}, Laura and {Hoenig}, Sebastian F. and {Imanishi}, Masatoshi and {Labiano}, Alvaro and {Levenson}, Nancy A. and {Pereira-Santaella}, Miguel and {Ramos Almeida}, Cristina and {Ricci}, Claudio and {Riffel}, Rogemar A. and {Rouan}, Daniel and {Rosario}, David and {Sandstrom}, Karin and {Shimizu}, Taro T. and {Stalevski}, Marko and {Thatte}, Niranjan and {Veenema}, Oscar and {Zhang}, Lulu},
        title = "{GATOS XIV: the first direct kinematic evidence of dusty outflows from AGN via PAH kinematics of local Seyfert galaxies with JWST}",
      journal = {\mnras},
         year = 2026,
        month = aug,
       volume = {550},
       number = {2},
          eid = {stag1181},
        pages = {stag1181},
          doi = {10.1093/mnras/stag1181},
archivePrefix = {arXiv},
       eprint = {2603.12200},
 primaryClass = {astro-ph.GA},
       adsurl = {https://ui.adsabs.harvard.edu/abs/2026MNRAS.550g1181D}
}

@article{das2006kinematics,
  title={Kinematics of the narrow-line region in the Seyfert 2 galaxy NGC 1068: dynamical effects of the radio jet},
  author={Das, Varendra and Crenshaw, DM and Kraemer, SB and Deo, RP},
  journal={The Astronomical Journal},
  volume={132},
  number={2},
  pages={620--632},
  year={2006}
}

@ARTICLE{emcee2013,
       author = {{Foreman-Mackey}, Daniel and {Hogg}, David W. and {Lang}, Dustin and {Goodman}, Jonathan},
        title = "{emcee: The MCMC Hammer}",
      journal = {\pasp},
         year = 2013,
        month = mar,
       volume = {125},
       number = {925},
        pages = {306},
          doi = {10.1086/670067},
archivePrefix = {arXiv},
       eprint = {1202.3665},
 primaryClass = {astro-ph.IM},
       adsurl = {https://ui.adsabs.harvard.edu/abs/2013PASP..125..306F}
}

@ARTICLE{Lai2022,
       author = {{Lai}, Thomas S.-Y. and {Armus}, Lee and {U}, Vivian and {D{\'\i}az-Santos}, Tanio and {Larson}, Kirsten L. and {Evans}, Aaron and {Malkan}, Matthew A. and {Appleton}, Philip and {Rich}, Jeff and {M{\"u}ller-S{\'a}nchez}, Francisco and {Inami}, Hanae and {Bohn}, Thomas and {McKinney}, Jed and {Finnerty}, Luke and {Law}, David R. and {Linden}, Sean T. and {Medling}, Anne M. and {Privon}, George C. and {Song}, Yiqing and {Stierwalt}, Sabrina and {van der Werf}, Paul P. and {Barcos-Mu{\~n}oz}, Loreto and {Smith}, J.~D.~T. and {Togi}, Aditya and {Aalto}, Susanne and {B{\"o}ker}, Torsten and {Charmandaris}, Vassilis and {Howell}, Justin and {Iwasawa}, Kazushi and {Kemper}, Francisca and {Mazzarella}, Joseph M. and {Murphy}, Eric J. and {Brown}, Michael J.~I. and {Hayward}, Christopher C. and {Marshall}, Jason and {Sanders}, David and {Surace}, Jason},
        title = "{GOALS-JWST: Tracing AGN Feedback on the Star-forming Interstellar Medium in NGC 7469}",
      journal = {\apjl},
         year = 2022,
        month = dec,
       volume = {941},
       number = {2},
          eid = {L36},
        pages = {L36},
          doi = {10.3847/2041-8213/ac9ebf},
archivePrefix = {arXiv},
       eprint = {2209.06741},
 primaryClass = {astro-ph.GA},
       adsurl = {https://ui.adsabs.harvard.edu/abs/2022ApJ...941L..36L}
}

@ARTICLE{Kirkpatrick2017,
       author = {{Kirkpatrick}, Allison and {Alberts}, Stacey and {Pope}, Alexandra and {Barro}, Guillermo and {Bonato}, Matteo and {Kocevski}, Dale D. and {P{\'e}rez-Gonz{\'a}lez}, Pablo and {Rieke}, George H. and {Rodr{\'\i}guez-Mu{\~n}oz}, Lucia and {Sajina}, Anna and {Grogin}, Norman A. and {Mantha}, Kameswara Bharadwaj and {Pandya}, Viraj and {Pforr}, Janine and {Salvato}, Mara and {Santini}, Paola},
        title = "{The AGN-Star Formation Connection: Future Prospects with JWST}",
      journal = {\apj},
         year = 2017,
        month = nov,
       volume = {849},
       number = {2},
          eid = {111},
        pages = {111},
          doi = {10.3847/1538-4357/aa911d},
archivePrefix = {arXiv},
       eprint = {1706.09056},
 primaryClass = {astro-ph.GA},
       adsurl = {https://ui.adsabs.harvard.edu/abs/2017ApJ...849..111K}
}

@ARTICLE{Feltre2016,
       author = {{Feltre}, A. and {Charlot}, S. and {Gutkin}, J.},
        title = "{Nuclear activity versus star formation: emission-line diagnostics at ultraviolet and optical wavelengths}",
      journal = {\mnras},
         year = 2016,
        month = mar,
       volume = {456},
       number = {3},
        pages = {3354-3374},
          doi = {10.1093/mnras/stv2794},
archivePrefix = {arXiv},
       eprint = {1511.08217},
 primaryClass = {astro-ph.GA},
       adsurl = {https://ui.adsabs.harvard.edu/abs/2016MNRAS.456.3354F}
}

@ARTICLE{DaviesRL2016,
       author = {{Davies}, Rebecca L. and {Groves}, Brent and {Kewley}, Lisa J. and {Dopita}, Michael A. and {Hampton}, Elise J. and {Shastri}, Prajval and {Scharw{\"a}chter}, Julia and {Sutherland}, Ralph and {Kharb}, Preeti and {Bhatt}, Harish and {Jin}, Chichuan and {Banfield}, Julie and {Zaw}, Ingyin and {James}, Bethan and {Juneau}, St{\'e}phanie and {Srivastava}, Shweta},
        title = "{Dissecting galaxies: spatial and spectral separation of emission excited by star formation and AGN activity}",
      journal = {\mnras},
         year = 2016,
        month = oct,
       volume = {462},
       number = {2},
        pages = {1616-1629},
          doi = {10.1093/mnras/stw1754},
archivePrefix = {arXiv},
       eprint = {1607.05731},
 primaryClass = {astro-ph.GA},
       adsurl = {https://ui.adsabs.harvard.edu/abs/2016MNRAS.462.1616D}
}

@ARTICLE{DAgostino2019,
       author = {{D'Agostino}, Joshua J. and {Kewley}, Lisa J. and {Groves}, Brent A. and {Medling}, Anne and {Dopita}, Michael A. and {Thomas}, Adam D.},
        title = "{A new diagnostic to separate line emission from star formation, shocks, and AGNs simultaneously in IFU data}",
      journal = {\mnras},
         year = 2019,
        month = may,
       volume = {485},
       number = {1},
        pages = {L38-L42},
          doi = {10.1093/mnrasl/slz028},
archivePrefix = {arXiv},
       eprint = {1902.10295},
 primaryClass = {astro-ph.GA},
       adsurl = {https://ui.adsabs.harvard.edu/abs/2019MNRAS.485L..38D}
}

@ARTICLE{Durre2018,
       author = {{Durr{\'e}}, Mark and {Mould}, Jeremy},
        title = "{The AGN Ionization Cones of NGC 5728. I. Excitation and Nuclear Structure}",
      journal = {\apj},
         year = 2018,
        month = nov,
       volume = {867},
       number = {2},
          eid = {149},
        pages = {149},
          doi = {10.3847/1538-4357/aae68e},
archivePrefix = {arXiv},
       eprint = {1810.03258},
 primaryClass = {astro-ph.GA},
       adsurl = {https://ui.adsabs.harvard.edu/abs/2018ApJ...867..149D}
}

@ARTICLE{Falcao2024,
       author = {{Trindade Falc{\~a}o}, Anna and {Fabbiano}, G. and {Elvis}, M. and {Paggi}, A. and {Maksym}, W.~P.},
        title = "{Deep Chandra Observations of NGC 5728. III. Probing the High-resolution X-Ray Morphology and Multiphase Interstellar Medium Interactions in the Circumnuclear Region}",
      journal = {\apj},
         year = 2024,
        month = dec,
       volume = {977},
       number = {2},
          eid = {275},
        pages = {275},
          doi = {10.3847/1538-4357/ad8de3},
archivePrefix = {arXiv},
       eprint = {2410.24061},
 primaryClass = {astro-ph.GA},
       adsurl = {https://ui.adsabs.harvard.edu/abs/2024ApJ...977..275T}
}

@ARTICLE{wilson1993,
       author = {{Wilson}, A.~S. and {Braatz}, J.~A. and {Heckman}, T.~M. and {Krolik}, J.~H. and {Miley}, G.~K.},
        title = "{The Ionization Cones in the Seyfert Galaxy NGC 5728}",
      journal = {\apjl},
         year = 1993,
        month = dec,
       volume = {419},
        pages = {L61},
          doi = {10.1086/187137},
       adsurl = {https://ui.adsabs.harvard.edu/abs/1993ApJ...419L..61W}
}

@ARTICLE{Shin2019,
       author = {{Shin}, Jaejin and {Woo}, Jong-Hak and {Chung}, Aeree and {Baek}, Junhyun and {Cho}, Kyuhyoun and {Kang}, Daeun and {Bae}, Hyun-Jin},
        title = "{Positive and Negative Feedback of AGN Outflows in NGC 5728}",
      journal = {\apj},
         year = 2019,
        month = aug,
       volume = {881},
       number = {2},
          eid = {147},
        pages = {147},
          doi = {10.3847/1538-4357/ab2e72},
archivePrefix = {arXiv},
       eprint = {1907.00982},
 primaryClass = {astro-ph.GA},
       adsurl = {https://ui.adsabs.harvard.edu/abs/2019ApJ...881..147S}
}

@ARTICLE{HermosaMunoz2026,
       author = {{Hermosa Mu{\~n}oz}, L. and {Gonz{\'a}lez Fern{\'a}ndez}, J.~R. and {Alonso-Herrero}, A. and {Garc{\'\i}a-Bernete}, I. and {Gonz{\'a}lez-Mart{\'\i}n}, O. and {Pereira-Santaella}, M. and {L{\'o}pez-Rodr{\'\i}guez}, E. and {Ramos Almeida}, C. and {Garc{\'\i}a-Burillo}, S. and {Zhang}, L. and {Audibert}, A. and {Bellocchi}, E. and {Combes}, F. and {D{\'\i}az-Santos}, T. and {Esparza-Arredondo}, D. and {Garc{\'\i}a-Lorenzo}, B. and {Garc{\'\i}a-Mar{\'\i}n}, M. and {Hicks}, E.~K.~S. and {Labiano}, {\'A}. and {Levenson}, N.~A. and {Mart{\'\i}nez-Paredes}, M. and {Packham}, C. and {Riffel}, R.~A. and {Rigopoulou}, D. and {Schneider}, J. and {Villar-Mart{\'\i}n}, M.},
        title = "{The Galaxy Activity, Torus, and Outflow Survey (GATOS): XII. Unveiling physical processes in local active galaxies. Unsupervised hierarchical clustering of JWST MIRI/MRS observations}",
      journal = {\aap},
         year = 2026,
        month = apr,
       volume = {708},
          eid = {A297},
        pages = {A297},
          doi = {10.1051/0004-6361/202557220},
       adsurl = {https://ui.adsabs.harvard.edu/abs/2026A&A...708A.297H}
}

@ARTICLE{Riffel2026,
       author = {{Riffel}, Rogemar A. and {Colina}, Luis and {Costa-Souza}, Jos{\'e} Henrique and {Mainieri}, Vincenzo and {Pereira Santaella}, Miguel and {Dors}, Oli L. and {Garc{\'\i}a-Bernete}, Ismael and {Alonso-Herrero}, Almudena and {Audibert}, Anelise and {Bellocchi}, Enrica and {Bunker}, Andrew J. and {Campbell}, Steph and {Combes}, Fran{\c{c}}oise and {Davies}, Richard I. and {D{\'\i}az-Santos}, Tanio and {Donnan}, Fergus R. and {Esposito}, Federico and {Garc{\'\i}a-Burillo}, Santiago and {Garc{\'\i}a-Lorenzo}, Bego{\~n}a and {Gonz{\'a}lez Mart{\'\i}n}, Omaira and {Haidar}, Houda and {Hicks}, Erin K.~S. and {Hoenig}, Sebastian F. and {Imanishi}, Masatoshi and {Labiano}, Alvaro and {Lopez-Rodriguez}, Enrique and {Packham}, Christopher and {Ramos Almeida}, Cristina and {Rigopoulou}, Dimitra and {Rosario}, David and {Souza-Oliveira}, Gabriel Luan and {Villar Mart{\'\i}n}, Montserrat and {Veenema}, Oscar and {Zhang}, Lulu},
        title = "{Impact of active galactic nuclei and nuclear star formation on the ISM turbulence of galaxies: Insights from JWST/MIRI spectroscopy}",
      journal = {\aap},
         year = 2026,
        month = jan,
       volume = {705},
          eid = {A59},
        pages = {A59},
          doi = {10.1051/0004-6361/202556775},
archivePrefix = {arXiv},
       eprint = {2510.02517},
 primaryClass = {astro-ph.GA},
       adsurl = {https://ui.adsabs.harvard.edu/abs/2026A&A...705A..59R}
}

@ARTICLE{GarciaBernete2024PAH,
       author = {{Garc{\'\i}a-Bernete}, I. and {Rigopoulou}, D. and {Donnan}, F.~R. and {Alonso-Herrero}, A. and {Pereira-Santaella}, M. and {Shimizu}, T. and {Davies}, R. and {Roche}, P.~F. and {Garc{\'\i}a-Burillo}, S. and {Labiano}, A. and {Hermosa Mu{\~n}oz}, L. and {Zhang}, L. and {Audibert}, A. and {Bellocchi}, E. and {Bunker}, A. and {Combes}, F. and {Delaney}, D. and {Esparza-Arredondo}, D. and {Gandhi}, P. and {Gonz{\'a}lez-Mart{\'\i}n}, O. and {H{\"o}nig}, S.~F. and {Imanishi}, M. and {Hicks}, E.~K.~S. and {Fuller}, L. and {Leist}, M. and {Levenson}, N.~A. and {Lopez-Rodriguez}, E. and {Packham}, C. and {Ramos Almeida}, C. and {Ricci}, C. and {Stalevski}, M. and {Villar Mart{\'\i}n}, M. and {Ward}, M.~J.},
        title = "{The Galaxy Activity, Torus, and Outflow Survey (GATOS): V. Unveiling PAH survival and resilience in the circumnuclear regions of AGNs with JWST}",
      journal = {\aap},
         year = 2024,
        month = nov,
       volume = {691},
          eid = {A162},
        pages = {A162},
          doi = {10.1051/0004-6361/202450086},
archivePrefix = {arXiv},
       eprint = {2409.05686},
 primaryClass = {astro-ph.GA},
       adsurl = {https://ui.adsabs.harvard.edu/abs/2024A&A...691A.162G}
}

@article{haidar2026gatos,
  title={GATOS XI: Excess dust heating in the Narrow Line Regions of nearby AGN revealed with JWST/MIRI},
  author={Haidar, Houda and Rosario, David J and Garc{\'\i}a-Bernete, Ismael and Alonso-Herrero, Almudena and Audibert, Anelise and Campbell, Steph and Harrison, Chris M and Costa, Tiago and Mu{\~n}oz, Laura Hermosa and Combes, Fran{\c{c}}oise and others},
  journal={Monthly Notices of the Royal Astronomical Society},
  pages={stag069},
  year={2026},
  publisher={Oxford University Press}
}

@article{lopez2025gatos,
  title={GATOS. VIII. On the Physical Origin of the Extended Mid-infrared Emission in Active Galactic Nuclei},
  author={Lopez-Rodriguez, Enrique and Ramos Almeida, Cristina and Pereira-Santaella, Miguel and Garc{\'\i}a-Bernete, Ismael and Nikutta, Robert and Alonso-Herrero, Almudena and Audibert, Anelise and Bellocchi, Enrica and Bunker, Andrew and Campbell, Steph and others},
  journal={The Astrophysical Journal},
  volume={994},
  number={2},
  pages={206},
  year={2025},
  publisher={The American Astronomical Society}
}

@ARTICLE{chown2025,
       author = {{Chown}, Ryan and {Leroy}, Adam K. and {Bolatto}, Alberto D. and {Chastenet}, J{\'e}r{\'e}my and {Glover}, Simon C.~O. and {Indebetouw}, R{\'e}my and {Koch}, Eric W. and {Donovan Meyer}, Jennifer and {Pingel}, Nickolas M. and {Rosolowsky}, Erik and {Sandstrom}, Karin and {Sutter}, Jessica and {Tarantino}, Elizabeth and {Bigiel}, Frank and {Boquien}, M{\'e}d{\'e}ric and {Chiang}, I.-Da and {Dale}, Daniel A. and {Dalcanton}, Julianne J. and {Egorov}, Oleg V. and {Eibensteiner}, Cosima and {Grasha}, Kathryn and {Hassani}, Hamid and {He}, Hao and {Kim}, Jaeyeon and {Meidt}, Sharon and {Pathak}, Debosmita and {Sarbadhicary}, Sumit K. and {Stanimirovic}, Snezana and {Villanueva}, Vicente and {Williams}, Thomas G.},
        title = "{Relationships between Polycyclic Aromatic Hydrocarbons, Small Dust Grains, H$_{2}$, and H I in Local Group Dwarf Galaxies NGC 6822 and WLM Using JWST, ALMA, and the VLA}",
      journal = {\apj},
         year = 2025,
        month = jul,
       volume = {987},
       number = {1},
          eid = {91},
        pages = {91},
          doi = {10.3847/1538-4357/add73a},
archivePrefix = {arXiv},
       eprint = {2504.08069},
 primaryClass = {astro-ph.GA},
       adsurl = {https://ui.adsabs.harvard.edu/abs/2025ApJ...987...91C}
}

@ARTICLE{AlonsoHerrero2012,
       author = {{Alonso-Herrero}, Almudena and {Pereira-Santaella}, Miguel and {Rieke}, George H. and {Rigopoulou}, Dimitra},
        title = "{Local Luminous Infrared Galaxies. II. Active Galactic Nucleus Activity from Spitzer/Infrared Spectrograph Spectra}",
      journal = {\apj},
         year = 2012,
        month = jan,
       volume = {744},
       number = {1},
          eid = {2},
        pages = {2},
          doi = {10.1088/0004-637X/744/1/2},
archivePrefix = {arXiv},
       eprint = {1109.1372},
 primaryClass = {astro-ph.CO},
       adsurl = {https://ui.adsabs.harvard.edu/abs/2012ApJ...744....2A}
}

@article{garcia2013searching,
  title={Searching double-peaked emission-line profiles in the spectra of galaxies through the symmetry of the cross-correlation function},
  author={Garc{\'\i}a-Lorenzo, B},
  journal={Monthly Notices of the Royal Astronomical Society},
  volume={429},
  number={4},
  pages={2903--2909},
  year={2013},
  publisher={Oxford University Press}
}

@article{davies2014starburst,
  title={Starburst--AGN mixing--II. Optically selected active galaxies},
  author={Davies, Rebecca L and Kewley, Lisa J and Ho, I-Ting and Dopita, Michael A},
  journal={Monthly Notices of the Royal Astronomical Society},
  volume={444},
  number={4},
  pages={3961--3974},
  year={2014},
  publisher={Oxford University Press}
}

@article{law2021sdss,
  title={SDSS-IV MaNGA: refining strong line diagnostic classifications using spatially resolved gas dynamics},
  author={Law, David R and Ji, Xihan and Belfiore, Francesco and Bershady, Matthew A and Cappellari, Michele and Westfall, Kyle B and Yan, Renbin and Bizyaev, Dmitry and Brownstein, Joel R and Drory, Niv and others},
  journal={The Astrophysical Journal},
  volume={915},
  number={1},
  pages={35},
  year={2021},
  publisher={The American Astronomical Society}
}

@article{prada1999counterrotating,
  title={A counterrotating central component in the barred galaxy NGC 5728},
  author={Prada, F and Guti{\'e}rrez, CM},
  journal={The Astrophysical Journal},
  volume={517},
  number={1},
  pages={123--129},
  year={1999}
}

@article{audibert2023jet,
  title={Jet-induced molecular gas excitation and turbulence in the Teacup},
  author={Audibert, A and Almeida, C Ramos and Garc{\'\i}a-Burillo, Santiago and Combes, F and Bischetti, Manuela and Meenakshi, Moun and Mukherjee, D and Bicknell, Geoffrey and Wagner, AY},
  journal={Astronomy \& Astrophysics},
  volume={671},
  pages={L12},
  year={2023},
  publisher={EDP Sciences}
}

@article{marconcini2026miracle,
  title={MIRACLE: III. JWST/MIRI expose the hidden role of the AGN outflow in NGC 1068},
  author={Marconcini, C and Marconi, A and Ceci, M and Feltre, A and Tart{\.e}nas, M and Zubovas, K and Lamperti, I and Cresci, G and Ulivi, L and Mannucci, F and others},
  journal={Astronomy \& Astrophysics},
  volume={712},
  pages={A94},
  year={2026},
  publisher={EDP Sciences}
}

@article{ceci2026miracle,
  title={MIRACLE: II. Unveiling the multiphase gas interplay in the circumnuclear region of NGC 1365 via multicloud modeling},
  author={Ceci, M and Marconcini, C and Marconi, A and Feltre, A and Lamperti, I and Belfiore, F and Bertola, E and Bracci, C and Carniani, S and Cataldi, E and others},
  journal={Astronomy \& Astrophysics},
  volume={707},
  pages={A376},
  year={2026},
  publisher={EDP Sciences}
}

\section*{Affiliations}
\noindent

$^{1}$Department of Physics, University of Oxford, Keble Road, Oxford, OX1 3RH, UK\\
$^{2}$School of Sciences, European University Cyprus, Diogenes street, Engomi, 1516 Nicosia, Cyprus\\
$^{3}$Max Planck Institute for extraterrestrial Physics, Giessenbachstrasse 1, 85748, Garching, Germany\\
$^{4}$Instituto de F\'{i}sica Fundamental, CSIC, Calle Serrano 123, 28006 Madrid, Spain\\
$^{5}$Centro de Astrobiolog\'{i}a (CAB), CSIC-INTA, Camino Bajo del 497 Castillo s/n, E-28692 Villanueva de la Ca{\~n}ada, Madrid, Spain\\
$^{6}$Instituto de Astrof\'{i}sica de Canarias, Calle V\'{i}a L\'{a}ctea, s/n, E-38205, La Laguna, Tenerife, Spain\\
$^{7}$Departamento de Astrof\'{i}sica, Universidad de La Laguna, E-28206, La Laguna, Tenerife, Spain\\
$^{8}$Departmento de F\'{i}sica de la Tierra y Astrof\'{i}sica, Fac. de CC F\'{i}sicas, Universidad Complutense de Madrid, E-28040 Madrid, Spain\\
$^{9}$Instituto de F\'isica de Part\'iculas y del Cosmos IPARCOS, Fac. CC. F\'isicas, Universidad Complutense de Madrid, 28040 Madrid, Spain\\
$^{10}$LUX, Observatoire de Paris, Coll\`ege de France, PSL University, CNRS, Sorbonne University, Paris, France\\
$^{11}$Institute of Astrophysics, Foundation for Research and Technology\mbox{--}Hellas (FORTH), Heraklion, GR\mbox{-}70013, Greece\\
$^{12}$Department of Astronomy \& Astrophysics, University of California, San Diego, La Jolla,
CA 92093, USA\\
$^{13}$Instituto de Radioastronomía y Astrofísica (IRyA-UNAM), 3-72 (Xangari), 8701, Morelia, Mexico\\
$^{14}$Observatorio Astron\'{o}mico Nacional (OAN-IGN)-Observatorio de Madrid, Alfonso XII, 3, 28014 Madrid, Spain\\
$^{15}$Instituto de Radioastronom\'ia y Astrof\'isica (IRyA), Universidad Nacional Autonoma de Mexico, Mexico\\
$^{16}$Departamento de F{\'i}sica, Universidad de Oviedo, Campus de Llamaquique, C/ Calvo Sotelo s/n, 33007 Oviedo, Spain\\
$^{17}$Department of Physics and Astronomy, University of Alaska Anchorage, Anchorage, AK 99508-4664, USA\\
$^{18}$School of Physics and Astronomy, University of Southampton, Southampton, SO17 1BJ, UK\\
$^{19}$National Astronomical Observatory of Japan, National Institutes of Natural Sciences (NINS), 2-21-1 Osawa, Mitaka, Tokyo, 181-8588, Japan\\
$^{20}$Department of Astronomy, School of Science, The Graduate University for Advanced Studies, SOKENDAI, Mitaka, Tokyo, 181-8588, Japan\\
$^{21}$Telespazio UK for the European Space Agency (ESA), ESAC, Camino Bajo del Castillo s/n, 28692 Villanueva de la Ca\~nada, Madrid, Spain\\
$^{22}$Space Telescope Science Institute, San Martin Drive, Baltimore, MD 21218, USA\\
$^{23}$Department of Physics \& Astronomy, University of South Carolina, Columbia, SC 29208, USA\\
$^{24}$Department of Astronomy, University of Geneva, ch. d'Ecogia 16, 1290, Versoix, Switzerland\\
$^{25}$Instituto de Estudios Astrof\'isicos, Facultad de Ingenier\'ia y Ciencias, Universidad Diego Portales, Av. Ej\'ercito Libertador 441, Santiago, Chile\\
$^{26}$Departamento de F\'isica, CCNE, Universidade Federal de Santa Maria, 97105-900 Santa Maria, RS, Brazil\\
$^{27}$Astronomical Observatory, Volgina 7, 11060 Belgrade, Serbia\\
$^{28}$Sterrenkundig Observatorium, Universiteit Gent, Krijgslaan 281-S9, Gent B-9000, Belgium\\






\bsp	
\label{lastpage}
\end{document}